\documentclass[pra,aps,twocolumn,showpacs]{revtex4}
\usepackage{graphicx,amsmath}
\usepackage{bbm}
\usepackage{txfonts}

\newcommand{\rp}[1]{(\ref{#1})}

\newcommand{\abs}[1]{\left|{#1}\right|}

\newcommand{\av}[1]{\left\langle #1 \right\rangle}

\newcommand{\br}[1]{\langle #1|}

\newcommand{\ke}[1]{|#1\rangle}

\newcommand{\kb}[2]{\ke{#1}\br{#2}}

\newcommand{\al}[1]{^{(#1)}}
\newcommand{\da}{^\dagger}

\newcommand{\pt}[1]{\left( #1 \right)}
\newcommand{\pq}[1]{\left[ #1 \right]}
\newcommand{\pg}[1]{\left\{ #1 \right\}}

\newcommand{\lpt}[1]{\left( #1 \right.}
\newcommand{\lpq}[1]{\left[ #1 \right.}

\newcommand{\rpt}[1]{\left. #1 \right)}
\newcommand{\rpq}[1]{\left. #1 \right]}

\newcommand{\ee}{{\rm e}}
\newcommand{\ii}{{\rm i}}
\newcommand{\dd}{{\rm d}}

\newcommand{\id}{\mathbbm{1}}

\newcommand{\nn}{{\nonumber}}

\newcommand{\ver}{{\bf r}}
\newcommand{\va}{{\bf a}}

\newcommand{\vu}{{\bf u}}
\newcommand{\vv}{{\bf v}}

\newcommand{\CC}{{\cal C}}

\newcommand{\EE}{{\cal E}}

\newcommand{\JJ}{{\cal J}}
\newcommand{\KK}{{\cal K}}

\newcommand{\LL}{{\cal L}}
\newcommand{\MM}{{\cal M}}
\newcommand{\NN}{{\cal N}}

\newcommand{\QQ}{{\cal Q}}

\newcommand{\TT}{{\cal T}}

\newcommand{\WW}{{\cal W}}

\newcommand{\YY}{{\cal Y}}
\newcommand{\ZZ}{{\cal Z}}

\begin{document}
\title{Non-Markovian dynamics and steady-state entanglement of cavity arrays in finite-bandwidth squeezed reservoirs}
\author{S. Zippilli$^{1}$ and F. Illuminati$^{1,2,3}$}
\affiliation{$^{1}$\mbox{Dipartimento di Ingegneria Industriale, Universit\`a degli Studi di Salerno, Via Giovanni Paolo II 132, I-84084 Fisciano (SA), Italy}
\\
$^{2}$\mbox{INFN, Sezione di Napoli, Gruppo Collegato di Salerno, I-84084 Fisciano (SA), Italy}
\\
$^{3}$\mbox{CNISM Unit\`a di Salerno, I-84084 Fisciano (SA), Italy
}}
\date{February 16, 2014}

\begin{abstract}
When two chains of quantum systems are driven at their ends by a two-mode squeezed reservoir, they approach a steady state characterized by the formation of many entangled pairs. Each pair is made of one element of the first and one of the second chain. This effect has been already predicted under the assumption of broadband squeezing. Here we investigate the situation of finite-bandwidth reservoirs. This is done by modeling the driving bath as the output field of a non-degenerate parametric oscillator.
The resulting non-Markovian dynamics is studied within the theoretical framework of cascade open quantum systems. It is shown that the formation of pair-entangled structures occurs as long as
the normal-mode splitting of the arrays does not overcome the squeezing bandwidth of the reservoir.
\end{abstract}
\maketitle

\section{Introduction}

\begin{figure}[!t]
\centering
\includegraphics[height=7cm]{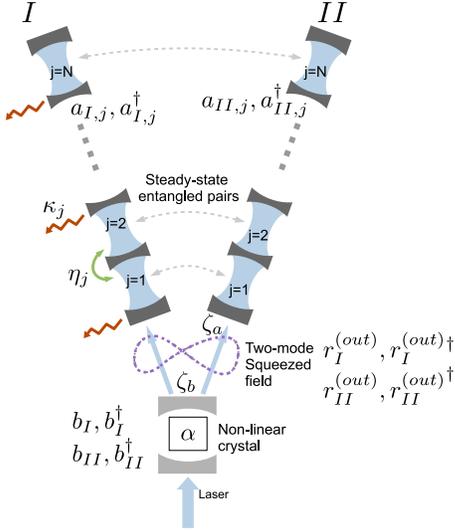}
\caption{Two arrays of cavities are driven at their ends by two-mode squeezed light. Arrows denote the inter-array pairs of cavities that become entangled at steady state when the squeezing bandwidth is sufficiently large.}
\label{schema}
\end{figure}

The ability to manipulate, engineer and control entanglement is of primary importance for the development of new quantum technologies, which are expected to outperform the corresponding classical implementations in many fields of application, including quantum computation, communication, metrology, and sensing.
In particular, the manipulation of entanglement over large quantum ensembles~\cite{Amico}, and the efficient transfer and distribution of quantum resources among different systems~\cite{Cirac97} are fundamental tasks for the development of scalable devices for quantum computation and information processing~\cite{Kimble}. In this context, a scheme was recently proposed~\cite{Zippilli2013} for the efficient transfer of quantum correlations from a broadband two-mode entangled reservoir to non-directly interacting chains of quantum systems. This phenomenon is observed at the steady state of the open incoherent dynamics in the framework of quantum reservoir engineering~\cite{Diehl,Verstraete}.
Specifically, it was shown~\cite{Zippilli2013}, that at the steady state, many pairs of entangled subsystems replicate the original entanglement of the reservoir, as illustrated in Fig.~\ref{schema}. This mechanism appears to be general, holding both for arrays of linearly coupled harmonic oscillators and two-level systems.

In the present work we generalize the previous investigation by relaxing the broadband condition of the reservoir. Indeed, for finite bandwidth the resulting dynamics of the arrays is in general non-Markovian. In order to study this problem we consider in detail the specific physical system producing the entangled reservoir. In this way, the full coupled dynamics of the arrays and of such system is Markovian and can be efficiently analyzed in the standard approach of cascade open quantum systems~\cite{Gardiner93,Gardiner94,Carmichael93,Carmichael94}. We will study the dynamics of two distinct arrays of linearly coupled harmonic oscillators interacting locally with the output field of a non-degenerate parametric oscillator. The latter is one of the most efficient sources of entangled fields and it is routinely used in many applications of quantum technologies~\cite{Neergaard-Nielsen,Lee,Steinlechner,Braunstein,Weedbrook}.
Its output is a continuous squeezed field whose statistical properties can be controlled through the characteristic parameters of the oscillator, essentially the strengths of the nonlinearity and of the dissipation.
We show that the pair-entanglement replication mechanism previously found in the case of a broadband reservoir continues to hold also in the more realistic scenario of finite bandwidth. This finding is thus
quite promising for the actual realizability of the mechanism in concrete experimental setups.

The paper is organized as follows. In Sec.~\ref{Model} we introduce the model in terms of quantum Langevin equations. In Sec.~\ref{Dynamics} we analyze the basic dynamics of the output field of the non-degenerate parametric oscillator (Sec.~\ref{Sec.PO}), and of the two arrays (Sec~\ref{Sec.arrays}). A relevant result of this analysis is the identification of the regimes in which one recovers the broadband Markovian dynamics considered previously~\cite{Zippilli2013}. In Sec~\ref{results}, we report  numerical results for the steady-state entanglement between the two arrays. In Sec.~\ref{Sec.Exp} we analyze the feasibility of the scheme with realistic experimental parameters. Finally, in Sec.~\ref{Conclusions} we draw our conclusions and discuss some possible outlooks.

\section{The Model}\label{Model}

We analyze the dynamics of two independent, non-directly interacting one-dimensional arrays (chains) of $N$ linearly coupled harmonic oscillators. In Fig.~\ref{schema} we illustrate a specific realization in terms of electromagnetic field modes confined inside cavities. One of the ends of each array is incoherently coupled, with rate $\zeta_a$, to the output field of a non-degenerate parametric oscillator. The coupling is non-reciprocal and realizes a cascade configuration~\cite{Gardiner93,Gardiner94,Carmichael93,Carmichael94,Gripp}, in which the arrays have no effect on the dynamics of the parametric oscillator. This situation can be realized using non-reciprocal devices such as, for instance, standard optical circulators based on the Faraday effect~\cite{Gripp,Stannigel}.
The creation and annihilation operators for the two arrays are $a_{\xi,j}$ and $a_{\xi,j}\da$, where the index $\xi$ takes the values $I$ or $II$ and distinguishes between the two arrays, while $j$ indicates the element in each array, as shown in Fig.~\ref{schema}. The oscillators in each chain are resonant, but the frequency of the two chains can be different. In each array, cavities are coupled with nearest-neighbor strength $\eta_j$ and lose excitations at a rate $\kappa_j$. The two arrays are assumed to be symmetric, i.e. they have the same interaction strengths $\eta_j$ and loss rates $\kappa_j$. Deviations from this symmetric condition have been previously investigated~\cite{Zippilli2013} showing that the replication mechanism is significantly resilient to random asymmetries. The total Hamiltonian for the two arrays, in interaction picture is
\begin{eqnarray}\label{Ha}
H_a&=&\sum_{\xi\in\pg{I,II}}\sum_{j=1}^{N-1}\eta_j\pt{{a_{\xi,j}}\da a_{\xi,j+1}+{a_{\xi,j+1}}\da a_{\xi,j}}\ .
\end{eqnarray}
The two modes of the nondegenerate parametric oscillator are $b_{\xi}$ and $b_{\xi}\da$, where $\xi\in\pg{I,II}$ indicates, respectively, the modes whose output drives the array $I$ and $II$, and they are resonant at the frequencies of the cavities of the corresponding arrays.
Moreover, they mutually interact with coupling strength $\alpha$. Throughout the present work the nonlinear coupling strength $\alpha$ is assumed, without loss of generality, to be real and positive. A different assumption would correspond to a different squeezing phase, but the overall entanglement dynamics will remain unaffected.
The corresponding parametric Hamiltonian reads
\begin{eqnarray}\label{Hb}
H_b&=&\ii\alpha\pt{ {b_I}\da {b_{II}}\da-b_I b_{II}}\ .
\end{eqnarray}
The two modes are further coupled to the output modes which drive the arrays with rate $\zeta_b$, see Fig.~\ref{schema}. The output field operators are $r_\xi\al{out}$ and ${r_\xi\al{out}}\da$ and they fulfill the standard bosonic field commutation relations $[r_\xi\al{out}(t),{r_{\xi'}\al{out}(t')}\da]=\delta_{\xi,\xi'}\delta(t-t')$.

The cascade quantum dynamics is conveniently described in terms of Heisenberg quantum Langevin equations for the field operators~\cite{Gardiner93,Gardiner94}. Then, the equations of motion for each array read
\begin{eqnarray}\label{QLEa}
\dot a_{\xi,1}(t)&=&-\pt{\kappa_1+\zeta_a}\ a_{\xi,1}(t)-\ii \eta_1\ a_{\xi,2}(t)
\\&&-\sqrt{2\kappa_1}\ a_{\xi,1}\al{in}(t)-\sqrt{2\zeta_a}\ r_{\xi}\al{out}(t),
\nn\\
\dot a_{\xi,j}(t)&=&-\kappa_j\ a_{\xi,j}(t)-\ii \eta_{j-1}\ a_{\xi,j-1}(t)-\ii \eta_{j}\ a_{\xi,j}(t),
\nn\\&&
-\sqrt{2\kappa_j}\ a_{\xi,1}\al{in}(t)
\hspace{1cm} {\rm for}\ j\in\pg{2...N-1}
\nn\\
\dot a_{\xi,N}(t)&=&-\kappa_N\ a_{\xi,N}(t)-\ii \eta_{N-1}\ a_{\xi,N-1}(t)-\sqrt{2\kappa_N}\ a_{\xi,N}\al{in}(t), \nn
\end{eqnarray}
with $\xi\in\pg{I,II}$. Here
$a_{\xi,j}\al{in}$ are the zero-average input noise operators for each cavity, whose only nonvanishing correlation function is $\av{a_{\xi,j}\al{in}(t){a_{\xi',j'}\al{in}}\da(t')} = \delta_{\xi,\xi'}\delta_{j,j'}\delta(t-t')$. We remark that in Eq.~\rp{QLEa} the operator for the output field of the nondegenerate parametric oscillator, $r_{\xi}\al{out}$, acts as a source term in the equation for the first cavity of each array.
The equations of motion for the field operators of the parametric oscillator are
\begin{eqnarray}\label{QLEb}
\dot b_I(t)&=&-(\kappa_0+\zeta_b)\ b_I(t)+\alpha\ b_{II}\da(t)
\nn\\&&
-\sqrt{2\kappa_{0}}\ b_I\al{in}(t)
-\sqrt{2\zeta_b}\ r_I\al{in}(t),
\nn\\
\dot b_{II}(t)&=&-(\kappa_0+\zeta_b)\ b_{II}(t)+\alpha\ b_{I}\da(t)
\nn\\&&
-\sqrt{2\kappa_{0}}\ b_{II}\al{in}(t)
- \sqrt{2\zeta_b}\ r_{II}\al{in}(t),
\end{eqnarray}
where $b_\xi\al{in}$ and $r_\xi\al{in}$ are the corresponding input noise operators, whose correlation functions satisfy $\av{b_{\xi}\al{in}(t){b_{\xi'}\al{in}}\da(t')}=\av{r_{\xi}\al{in}(t){r_{\xi'}\al{in}}\da(t')}
= \delta_{\xi,\xi'}\delta(t-t')$.
The noise operator $r_\xi\al{in}$ accounts for the modes of the reservoir which are involved in the driving process. The operator $b_\xi\al{in}$ accounts for the other possible dissipation channels, with decay rate $\kappa_0$, whose output field is lost and cannot be used to drive the two arrays.
The output field operators corresponding to each input operator are related to the system operators (parametric oscillator and arrays) through the standard relations~\cite{Gardiner93,Gardiner94}
\begin{eqnarray}\label{rout}
r_\xi\al{out}(t)&=&r_\xi\al{in}(t)+\sqrt{2\zeta_b}\ b_\xi(t),
\\
a_{\xi,j}\al{out}(t)&=&a_{\xi,j}\al{in}(t)+\sqrt{2\kappa_j}a_{\xi,j}(t),
\\
b_{\xi}\al{out}(t)&=&b_{\xi}\al{in}(t)+\sqrt{2\kappa_0}b_{\xi}(t).
\end{eqnarray}
In particular Eq.~\rp{rout} can be used to eliminate the output field in the equations for the arrays and to
make explicit their dependence on the operators of the parametric oscillator and on the corresponding input noise field.

\section{Dynamics} \label{Dynamics}

The dynamics described by the linear equations~\rp{QLEa}, \rp{QLEb} and \rp{rout} is Gaussian,
hence it is completely determined by the equations for the average and the correlations of the field operators (first and second statistical moments). In particular, assuming that the initial fields averages are zero, they will remain zero at all times. This is the situation that we assume through the rest of this article. Hence, in the following, we will focus our investigation on the analysis of the correlation matrix of the array operators $\CC_a(t)$, whose elements are defined, in terms of the vector of the $4N$ field operators $\va=\pt{a_{I,1}\cdots a_{I,N},a_{II,1}\cdots a_{II,N},a_{I,1}\da\cdots a_{I,N}\da,a_{II,1}\da\cdots a_{II,N}\da}$, as
\begin{eqnarray}\label{Cat}
\pg{\CC_a(t)}_{j,k}=\av{\pg{\va(t)}_j\ \pg{\va(t)}_k}\ .
\end{eqnarray}
This matrix contains all the information about the state of the arrays and can be used, for example, to analyze the entanglement properties of the steady state. In particular, we will be interested in the bipartite entanglement of pairs of inter-array cavity modes, quantified by the logarithmic negativity.

The logarithmic negativity,  $\EE_N$, for two generic bosonic modes can be computed as follows. Given a generic correlation matrix $\CC$ for two modes
the corresponding logarithmic negativity $\EE_N$ can be easily evaluated as~\cite{Adesso}:
\begin{eqnarray}\label{EN}
\EE_N={\rm max}\pg{0,-\log\nu_-},
\end{eqnarray}
where $\nu_-$ denotes the smallest symplectic eigenvalue of the matrix $\frac{1}{2}\TT(\CC+\CC^T)\TT^T$
with
\begin{eqnarray}
\TT=\pt{
\begin{array}{cccc}
 1 &  0 & 1 & 0 \\
 \ii &  0 & -\ii & 0\\
0  & 1  & 0 & 1 \\
0& -\ii &0 & \ii
\end{array}
}\ .
\end{eqnarray}
In the following sections we will use this definition to compute both the logarithmic negativity of two cavity modes $\EE_N\al{cav}[j_I,j_{II}]$, where $j_I$ and $j_{II}$ are the cavity indices of the first and second array respectively, and  the logarithmic negativity of the two modes of the reservoir $\EE_N\al{PO}$.

\subsection{
Spectral properties of the output field of the nondegenerate parametric oscillator
}
\label{Sec.PO}

Before considering in detail the results for the steady state of the open dynamics for the two arrays,
in this subsection we describe the main features of the steady state of the nondegenerate parametric oscillator which are useful in the following. The open dynamics of a nondegenerate parametric oscillator (see~\cite{Reid1,Reid2,Reid3,Reid4} for a detailed discussion) is described by the quantum Langevin equation \rp{QLEb} which
admits a steady state solution only if the rates for the decay processes are sufficiently large to counterbalance the effects of the parametric amplification. In this case the parametric oscillator is said to work below threshold. The condition for this regime is
$$\alpha<\zeta_b+\kappa_0 \, ,$$
where $\zeta_b+\kappa_0$ accounts for the total dissipation rate. This condition will always be assumed in the following.

The output field is completely characterized by the two-time correlation matrix $\CC\al{out}_r(t,t')$
whose elements are defined, in terms of the elements of the vector of the four output field operators $\ver_{out}=\pt{r\al{out}_I,r\al{out}_{II},{r\al{out}_I}\da,{r\al{out}_{II}}\da }$, as
\begin{eqnarray}\label{Coutrtt}
\pg{\CC\al{out}_r(t,t')}_{j,k}=\av{\pg{\ver_{out}(t)}_j\ \pg{\ver_{out}(t')}_k}\ .
\end{eqnarray}
The corresponding steady-state correlation matrix $\CC\al{out}_{r,st}(t-t')$ depends only on the difference of the time arguments, and
can be expressed as
\begin{eqnarray}\label{Coutr}
\CC\al{out}_{r,st}(\tau)
&=&\delta(\tau)\ \YY+\frac{\alpha\ \zeta_b}{2}\sum_{\iota=\pm} \frac{\ee^{-\bar\alpha_\iota\ \abs{\tau}}}{\bar\alpha_\iota}\WW_\iota
\end{eqnarray}
where the parameters $\bar \alpha_\pm$ are the decay rates of the field correlation functions. In turn, as discussed below, they determine the bandwidth over which the statistical properties of the reservoir are of the same order of magnitude. They are defined as
\begin{eqnarray}\label{baralpha+-}
\bar\alpha_\pm=\zeta_b+\kappa_0\pm\alpha\ .
\end{eqnarray}
In Eq.~\rp{Coutr}, the $4\times4$ matrix $\YY$ has only two non-zero elements:
\begin{eqnarray}\label{Y}
\pg{\YY}_{1,3}=\pg{\YY}_{2,4}=1\ .
\end{eqnarray}
It corresponds to the correlation matrix for two modes in vacuum, and it is asymmetric and nonvanishing because of the non-commutativity of the corresponding field operators.
Moreover, $\WW_\pm$ is the $4\times 4$ matrix
\begin{eqnarray}\label{E}
\WW_\pm=\pt{
\begin{array}{cccc}
 0 &  1 & \mp1 & 0 \\
 1 &  0 & 0 & \mp1 \\
\mp1  & 0  & 0 & 1 \\
0&\mp1&1&0
\end{array}
}\ ,
\end{eqnarray}
and the part of Eq.~\rp{Coutr} that contains these matrices accounts for a finite number of excitations and for the correlations of the two modes.
The corresponding spectral density matrix $\widetilde\CC_r\al{out}(\omega)=\int_{-\infty}^\infty\dd \tau\ \ee^{\ii\omega \tau}\ \CC_{r,st}\al{out}(\tau)$ is given by
\begin{eqnarray}\label{CCoutom}
\widetilde\CC_r\al{out}(\omega)&=&
\YY+\sum_{\iota=\pm} \frac{\alpha\ \zeta_a}{\bar\alpha_\iota^2+\omega^2}\WW_\iota\ .
\end{eqnarray}
It is the sum of two Lorentzian whose bandwidths are $\bar \alpha_+$ and $\bar \alpha_-$ respectively,
and it is related to the correlation functions of the output field in frequency space
and, hence, to the spectrum of the squeezed reservoir
as follows. Let as define the vector of operators in frequency space $\tilde\ver_{out}(\omega)=\pt{\tilde r\al{out}_I(\omega),\tilde r\al{out}_{II}(\omega),{\tilde r\al{out}_I}{}\da(\omega),{\tilde r\al{out}_{II}}{}\da(\omega) }$, with $\tilde r_\xi\al{out}(\omega)=\frac{1}{\sqrt 2}\int_{-\infty}^\infty\dd \tau\ \ee^{\ii\omega \tau}\ r_\xi\al{out}(\tau)$, which implies that $\pg{\tilde r_\xi\al{out}(\omega)}\da={\tilde r_\xi\al{out}}{}\da(-\omega)$. Then $\av{\pg{\tilde\ver_{out}(\omega)}_j\ \pg{\tilde\ver_{out}(\omega')}_k}=\delta(\omega+\omega')\ \widetilde\CC_r\al{out}(\omega)$.
We note that, since the system Hamiltonian is defined in interaction picture, the frequency $\omega$ is not an absolute frequency, but it is defined relative to the frequencies of the modes of the nondegenerate parametric oscillator, which in turn are equal to the frequencies of the cavity modes. Being relative, it can take both positive and negative values.
In particular the mode at $\omega=0$ in interaction picture, $\tilde r_\xi\al{out}(0)$, corresponds to the spectral component of the output field at the frequency of the cavities of the array $\xi$ in the original picture.

The spectral density matrix in Eq.~\rp{CCoutom} can be used to construct the
squeezing spectrum, that is the noise spectral density associated to the collective quadrature $X=\pt{r_I\al{out}+{r_I\al{out}}\da-r_{II}\al{out}-{r_{II}\al{out}}\da}/\sqrt{2}\equiv \vu^T\ \ver_{out}$,  where $\vu=(1,-1,1,-1)^T$. It is given by
\begin{eqnarray}\label{Sq}
S(\omega)&=&\vu^T\ \widetilde\CC_r\al{out}(\omega)\ \vu
\nn\\
&=&1-\frac{4\ \alpha\ \zeta_b}{\bar \alpha_+^2+\omega^2}.
\end{eqnarray}
According to our definitions, the output field is two-mode squeezed when $S(\omega)<1$.
Specifically, Eq.~\rp{Sq} indicates that the spectral components of the output field are squeezed over a bandwidth  $\bar \alpha_+$, with maximum squeezing attained at the central frequency $\omega=0$.
On the other hand, the orthogonal quadrature $Y=\vv^T\ \ver_{out}$,  with $\vv=\ii(1,-1,-1,1)^T$,
is anti-squeezed, namely its variance is larger then one. Indeed, the corresponding noise spectrum
is given by
\begin{eqnarray}\label{aSq}
T(\omega)=1+\frac{4\ \alpha\ \zeta_b}{\bar \alpha_-^2+\omega^2} \, ,
\end{eqnarray}
which exhibits anti-squeezing over a bandwidth $\bar\alpha_-$.
It is interesting to point out that squeezed and anti-squeezed quadratures are characterized by non-equal bandwidths, which can be even quite different. In particular, this is true when the squeezing is maximum.
Indeed, maximum squeezing is obtained in the limit $\bar\alpha_-\to 0$, namely at the threshold of the parametric oscillator. In this case, perfect suppression of the fluctuations of the squeezed quadrature
is observed at the central frequency $\omega=0$, that is $S(0)\sim 0$; then, as already remarked, the squeezing extends over a finite bandwidth $\bar\alpha_+$. On the other hand, in order to satisfy the Heisenberg uncertainty relation, the fluctuations of the orthogonal anti-squeezed quadrature must diverge at the central frequency and the corresponding bandwidth vanishes.

Two-mode squeezing is a strong sufficient condition for entanglement in continuous-variable systems~\cite{Braunstein,Weedbrook}. In particular, the squeezing spectrum at a given frequency $\omega$ measures the entanglement between the spectral component at frequency $\omega$ of the first mode,
and the spectral component at frequency $-\omega$ of the second one.
The corresponding logarithmic negativity
$\EE_N\al{PO}(\omega)$ can be evaluated applying the definition in Eq.~\rp{EN} to the correlation matrix $\widetilde\CC_r\al{out}(\omega)$ defined in Eq.~\rp{CCoutom}.
In our case, the smallest symplectic eigenvalue $\nu_-(\omega)$,
evaluated for each spectral component,
is indeed equal to the value of the corresponding squeezing spectrum
\begin{eqnarray}
\nu_-(\omega)=S(\omega).
\end{eqnarray}
Correspondingly, maximum entanglement is found for the spectral components at the central frequency $\omega=0$, and the entanglement extends over a bandwidth $\bar\alpha_+$.

\subsection{Reduced non-Markovian dynamics for the two arrays}\label{Sec.arrays}

To gain insight into the system dynamics and in order to draw a clear comparison with the case of a broadband squeezed reservoir~\cite{Zippilli2013}, in this subsection we discuss the reduced non-Markovian dynamics of the two arrays that is obtained after tracing out the degrees of freedom of the parametric oscillator.
We anticipate that no approximation is performed in the derivation of the reduced dynamical equations, hence the corresponding dynamics of the arrays is equal to that obtained directly from the full model in Eqs.~\rp{QLEa}, \rp{QLEb} and \rp{rout}.

The equations~\rp{QLEb} and \rp{rout} for the nondegenerate parametric oscillator do not depend on the degrees of freedom of the arrays. Therefore, they can be solved and used to obtain a set of closed equations for the dynamics of the arrays alone. The resulting equation for the correlation matrix of the arrays $\CC_a(t)$, whose definition was given in Eq.~\rp{Cat}, is linear with a time-dependent source term of the form
\begin{eqnarray}\label{dotCCa}
\frac{\partial}{\partial t}\CC_a(t)=\MM_a\ \CC_a(t)+\CC_a(t)\ \MM_a^T+\NN_a(t),
\end{eqnarray}
where
$\MM_a$ is the $4N\times 4N$ matrix of coefficients corresponding to
the homogeneous part of
the system of equations~\rp{QLEa}; it is a tridiagonal matrix
whose elements, along the main diagonal, are
\begin{eqnarray}\label{Ma_d0}
&&\pg{\MM_a}_{1+\ell,1+\ell}=-(\zeta_a+\kappa_1)\ \ \ {\rm for}\ \ell\in\pg{0,N,2N,3N}\ ,
\nn\\
&&\pg{\MM_a}_{j+\ell,j+\ell}=-\kappa_j\ \ {\rm for}\ \ell\in\pg{0,N,2N,3N},\  {\rm and}\ j\in\pg{2\cdots N},
\nn\\
\end{eqnarray}
while the non-zero elements along the first upper and lower diagonals are
\begin{eqnarray}\label{Ma_d1}
&&\pg{\MM_a}_{j+\ell,j+\ell+1}=\pg{\MM_a}_{j+\ell+1,j+\ell}=-\pg{\MM_a}_{j+\ell+2N,j+\ell+2N+1}
\nn\\&&\hspace{2.5cm}
=-\pg{\MM_a}_{j+\ell+2N+1,j+\ell+2N}=-\ii\eta_j
\nn\\&&\hspace{3cm}
{\rm for}\ \ell\in\pg{0,N}, \ {\rm and}\ j\in\pg{1\cdots N-1}.
\nn\\
\end{eqnarray}
The time-dependent inhomogeneous source term $\NN_a(t)$ in Eq.~\rp{dotCCa} is a $4N\times4N$ matrix defined as
\begin{eqnarray}\label{Nat0}
\NN_a(t)&=&\QQ_a+2\zeta_a\int_0^t\dd\tau\ \lpq{
\ee^{\MM_a (t-\tau)}
{\ZZ}\ \CC\al{out}_r(\tau,t)\
{\ZZ}^T
}\nn\\&&\hspace{-1cm} \rpq{
+\ {\ZZ}\ \CC\al{out}_r(t,\tau)\
{\ZZ}^T
\ee^{{\MM_a}^T (t-\tau)},
}
\end{eqnarray}
where $\QQ_a$ is a matrix whose non-zero elements are
$\pg{\QQ_a}_{j+\ell,2N+j+\ell}=2\kappa_j$ for $\ell\in\pg{0,N}$ and $j\in\pg{1\cdots N}$.
The quantity $\CC\al{out}_r(t,t')$ is the $4\times4$ correlation matrix already defined in Eq.~\rp{Coutrtt}.
Finally, $\ZZ$ is the $4N\times 4$ matrix defined as
\begin{eqnarray}\label{ZZ}
\ZZ=\pt{
\begin{array}{cccc}
1          &             &   & \\
 \vdots &    & & \\
 0          &            &  &  \\
           & 1         &  &  \\
           & \vdots  &   & \\
          &  0       &   & \\
             &           & 1          &\\
             &           & \vdots  &\\
             &           & 0          &\\
             &           &            &1\\
             &           &            &\vdots\\
             &           &            &0\\
\end{array}
} \, ,
\end{eqnarray}
where the missing entries are all zeros.

In particular, if we assume that the parametric oscillator is, at the initial time of the arrays dynamics, in its stationary state, then
the result in Eq.~\rp{Coutr} can be used to express $\NN_a(t)$ as
\begin{eqnarray}\label{NNa}
&&\NN_a(t)=\QQ_a+2\zeta_a\ \ZZ\YY\ZZ^T
-\alpha\ \zeta_b\ \zeta_a\sum_{\iota=\pg{+,-}}\frac{1}{\bar\alpha_\iota}
\\&&\hspace{.5cm}\times
\pq{
\frac{1-\ee^{\pt{\MM_a-\bar\alpha_\iota}t}}{\MM_a-\bar\alpha_\iota} \ \ZZ\WW_\iota\ZZ^T
+ \ZZ\WW_\iota\ZZ^T\ \frac{1-\ee^{\pt{\MM_a^T-\bar\alpha_\iota}t}}{\MM_a^T-\bar\alpha_\iota}
},\nn
\end{eqnarray}
where $\YY$ and $\WW_\iota$ are defined in Eqs.~\rp{Y} and \rp{E} respectively.
The first two terms of this expression account for the dissipations of the arrays, while the last term is due to the effect of the entangled reservoir. It depends on both the properties of the output field of the parametric oscillator (through the parameters $\bar\alpha_\pm$) and of the two arrays (through the matrix $\MM_a$).

In the following, we are interested in the steady state of the arrays. The corresponding correlation matrix $\CC_a\al{st}$ can be computed from Eq.~\rp{dotCCa}.
Defining the linear operator $\LL_a$ which operates on a generic correlation matrix $\CC$ as $\LL_a\CC=\MM_a\ \CC+\CC\ \MM_a^T$, the steady state can then be expressed formally as
\begin{eqnarray}\label{Cast}
\CC_a\al{st}=-\LL_a^{-1}\ \NN_0\ ,
\end{eqnarray}
where $\NN_0$ is the time-independent part of Eq.~\rp{NNa}, namely
\begin{eqnarray}
&&\NN_0=\QQ_a+2\zeta_a\ \ZZ\YY\ZZ^T
-\alpha\ \zeta_b\ \zeta_a\sum_{\iota=\pg{+,-}}\frac{1}{\bar\alpha_\iota}
\\&&\hspace{1.5cm}\times
\pq{
\frac{1}{\MM_a-\bar\alpha_\iota} \ \ZZ\WW_\iota\ZZ^T
+ \ZZ\WW_\iota\ZZ^T\ \frac{1}{\MM_a^T-\bar\alpha_\iota}
}.\nn
\end{eqnarray}

We finally note that Eq.~\rp{dotCCa} is equivalent to the following non-Markovian master equation in Lindblad form for the reduced density matrix $\rho_a$ of the two arrays:
\begin{eqnarray}\label{Meq_a}
\dot{\rho_a}&=&-\ii\pq{H_a\ ,\ \rho_a}
\\
&&+\sum_{j,k} \pg{\KK_a(t)}_{j,k} \pq{2 \va_k\ \rho_a\ \va_j-\va_j\ \va_k \ \rho_a-\rho_a\ \va_j\ \va_k} \, ,
\nn
\end{eqnarray}
where $H_a$ is the Hamiltonian of the two arrays, Eq. \rp{Ha}, and the time-dependent Kossakowski matrix $\KK_a(t)$, responsible for the non-Markovian character of the dynamics, is defined as
\begin{eqnarray}
\KK_a(t)&=&\frac{1}{2} \JJ\ \NN_a(t)\ \JJ\ ,
\end{eqnarray}
with $\JJ$ the $4N\times4N$ symplectic matrix
\begin{eqnarray}\label{J}
\JJ=\pt{
\begin{array}{cc}
\  & \id_{2N} \\
-\id_{2N}\  &\
\end{array}
}\ ,
\end{eqnarray}
where $\id_{2N}$ is the $2N\times2N$ identity matrix, and the missing blocks are null matrices.

\subsubsection{Broadband limit}\label{Sec.BroadBand}

Eqs.~\rp{dotCCa} and \rp{Meq_a} allow for a direct analysis of the limit of infinite bandwidth. This corresponds to the situation in which the parameters $\bar\alpha_\pm$, defined in Eq.~\rp{baralpha+-}, are the largest in the system dynamics. Since, by definition, $\bar\alpha_+>\bar\alpha_-$, this limit is obtained when $\bar\alpha_-$ is much larger then the eigenvalues of $\MM_a$, that is, in particular, when
\begin{eqnarray}\label{broad_condition}
\bar\alpha_-\gg \zeta_a,\kappa_j,\eta_j.
\end{eqnarray}
In this limit, the steady state correlation matrix of the reservoir, Eq.~\rp{Coutr}, takes the form
\begin{eqnarray}\label{CoutrBroadband}
\CC\al{out}_{r,st}(\tau)
&=& \delta(\tau)\pt{\YY+\alpha\ \zeta_b\sum_{\iota=\pm} \frac{\WW_\iota}{\bar\alpha_\iota^2}}
\end{eqnarray}
and Eq.~\rp{NNa} reduces to
\begin{eqnarray}
&&\NN_a(t) = \QQ_a+2\zeta_a\ \ZZ\YY\ZZ^T
+2\alpha\ \zeta_b\ \zeta_a\sum_{\iota=\pg{+,-}}
\frac{\ZZ\WW_\iota\ZZ^T}{\bar\alpha_\iota^2}.
\nn\\
\end{eqnarray}
Hence the corresponding master equation for the reduced density matrix of the arrays is equal to the one introduced in~\cite{Zippilli2013}:
\begin{eqnarray}\label{MEq0}
\dot\rho_a&=&-\ii\pq{H_a,\rho_a}
\nn\\&&
+\sum_{j=1}^N\kappa_j\sum_{\xi=I,II} \pt{2 a_{\xi,j}\rho_a a_{\xi,j}\da-a_{\xi,j}\da a_{\xi,j}\rho_a -
\rho_a a_{\xi,j}\da a_{\xi,j}}
\nn\\&&
+\zeta_a\pt{\bar n+1}\sum_{\xi=I,II} \pt{2 a_{\xi,1}\rho_a a_{\xi,1}\da-a_{\xi,1}\da a_{\xi,1}\rho_a -
\rho_a a_{\xi,1}\da a_{\xi,1}}
\nn\\&&
+\zeta_a\ \bar n\sum_{\xi=I,II} \pt{2 a_{\xi,1}\da\rho_a a_{\xi,1}-a_{\xi,1} a_{\xi,1}\da\rho_a -
\rho_a a_{\xi,1} a_{\xi,1}\da}
\nn\\&&
-2\zeta_a\ \bar m \lpt{a_{I,1}\rho_a a_{II,1}+a_{II,1}\rho_a a_{I,1}
}\nn\\&&\rpt{
-a_{I,1} a_{II,1}\rho_a - \rho_a a_{I,1} a_{II,1}+{\rm h.c.}} \, ,
\end{eqnarray}
where the parameters $\bar n$ and $\bar m$ account, respectively, for the number of excitations and for the correlations of the reservoir, and are given by
\begin{eqnarray}
\bar n=\alpha\ \zeta_b\pt{\frac{1}{\bar\alpha_-^{2}}-\frac{1}{\bar\alpha_+^{2}}}, \ \ \ \ \ \ \ \ \bar m=\alpha\ \zeta_b\pt{\frac{1}{\bar\alpha_-^{2}}+\frac{1}{\bar\alpha_+^{2}}}\ .
\end{eqnarray}
Specifically, according to Eq.~\rp{CoutrBroadband}, they fulfill the relations
\begin{eqnarray}
\av{{r_\xi\al{out}}\da(t)\ r_\xi\al{out}(t')}_{st}&=&\bar n\ \delta(t-t'),\nn\\
\av{{r_I\al{out}}(t)\ r_{II}\al{out}(t')}_{st}&=&\bar m\ \delta(t-t'),
\end{eqnarray}
where the label $st$ indicates that the averages are performed over the steady state.

Therefore, in this limit we recover all the results of~\cite{Zippilli2013} and, in particular,
when $\kappa_j=0$, the exact steady state solution of Eq.~\rp{MEq0} takes the form
\begin{eqnarray}
\rho_a\al{st}=\bigotimes_{j=1}^N U_j\ \varrho_{I,j}\otimes\varrho_{II,j}\ U_j\da
\end{eqnarray}
where $\varrho_{\xi,j}$ is the thermal state
$$\varrho_{\xi,j}=\sum_{n=0}^{\infty}\frac{1}{1+\bar n_T}\pt{\frac{\bar n_T}{1+\bar n_T}}^n\kb{n}{n}\ ,$$
with $\bar n_T=\frac{1}{2}\pq{\sqrt{\pt{2\bar n+1}^2-4\bar m^2}-1}$ average photons, and it is the same for all the cavities.
Moreover
$U_j$ is a squeezing transformation for the two modes $a_{I,j}$ and $a_{II,j}$, and it is defined as
$$U_j=\ee^{(-1)^j s_0\ \pt{a_{I,j}a_{II,j}-a_{I,j}\da a_{II,j}\da}},$$
where the squeezing parameter $s_0$ can be expressed
as
\begin{eqnarray}
\tanh\pt{s_0}=\frac{\bar n-\bar n_T}{\bar m}\ .
\end{eqnarray}
Correspondingly, in the steady state, the correlation functions for the arrays are
\begin{eqnarray}
&&\av{a_{\xi,j}\da\ a_{\xi',j'}}_{st}=\bar n\ \delta_{j,j'}\ \delta_{\xi,\xi'}
\nn\\
&&\av{a_{\xi,j}\ a_{\xi',j'}\da}_{st}=(\bar n+1)\ \delta_{j,j'}\ \delta_{\xi,\xi'}
\nn\\
&&\av{a_{I,j}\ a_{II,j'}}_{st}=\av{a_{I,j}\da\ a_{II,j'}\da}_{st}=\pt{-1}^{j+1}\bar m\ \delta_{j,j'}
\nn\\
&&\av{a_{\xi,j}\ a_{\xi,j'}}_{st}=\av{a_{\xi,j}\da\ a_{\xi,j'}\da}_{st}=0\ ,
\end{eqnarray}
and, hence, the logarithmic negativity for each pair of cavities with equal indices along the two arrays is given by $\EE_N\al{cav}[j,j]=-\log\pt{2\bar n+1-2\bar m}=-\log\pt{1-4\alpha\zeta_b/\bar\alpha_+}$, which is equal to the logarithmic negativity between the modes of the reservoir.

In the next section we analyze numerically the validity of this limit, and how the results of~\cite{Zippilli2013} are modified when  condition~\rp{broad_condition} is not fulfilled.

A final remark is in order.
Eq.~\rp{broad_condition} implies that if the parametric oscillator is very close to the threshold condition ($\bar\alpha_-\sim0$), that is, as discussed in Sec.~\ref{Sec.PO}, when the entanglement is maximum, then the broadband assumption is no longer valid even if the squeezing spectrum is relatively broad (its bandwidth is, in fact, given by $\bar \alpha_+$).

\section{Results}
\label{results}

In this section we analyze numerically the entanglement properties of the steady state of the
two arrays.
We characterize the steady state in terms of the logarithmic negativity
$\EE_N\al{cav}[j_I,j_{II}]$ for the modes $j_I$ and $j_{II}$ of first and second array respectivelly,
which is evaluated applying the general definition,
Eq.~\rp{EN}, to the reduced correlation matrix of each pair of cavity modes which, in turn, is obtained form the full steady state correlation matrix defined in Eq.~\rp{Cast}. In particular, we analyze the normalized logarithmic negativity $E\al{cav}_N[j_I,j_{II}]={\EE_N\al{cav}[j_I,j_{II}]}/(\EE_N\al{cav}[j_I,j_{II}]+1)$,
which takes values between 0 and 1.

We are interested in identifying parameter regimes in which the entanglement replication mechanism described in~\cite{Zippilli2013} is still visible even if the assumptions assumed in~\cite{Zippilli2013} are not strictly satisfied. According to the mechanism of entanglement replication described in~\cite{Zippilli2013}, at the steady state the cavities with equal indices along the two arrays are entangled with their
logarithmic negativity equal to that of the driving two-mode squeezed field. This result is exact if the driving entangled reservoir is broadband. As discussed in Sec.~\ref{Sec.BroadBand} this limit is expressed by the condition Eq.~\rp{broad_condition}.

\begin{figure}[!t]
\includegraphics[width=8cm]{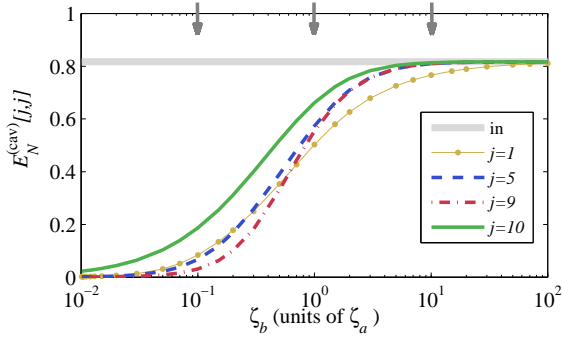}
\caption{Logarithmic negativity for inter-array pairs of cavities with equal indices along two equal arrays of $N=10$ cavities each, as a function of the emission rate $\zeta_b$ of the parametric oscillator, when the ratio $\alpha/\zeta_b=0.648$ is kept constant in order to fix the value of the entanglement at the central frequency of the output field, i.e. the reservoir that drives the two arrays. The inset specifies the
index of the cavities
corresponding to each curve. The gray solid thick line indicates the value of the entanglement at the central frequency of the reservoir.
The other parameters are, $\eta_j=\zeta_a$, $\forall j\in\pg{1,...N}$ and $\kappa_j=0$, $\forall j\in\pg{0,...N}$. The arrows on the top part of the plot indicate the parameters corresponding, respectively, to the results of Figs.~\ref{fig2_a}, \ref{fig2_b} and \ref{fig2_d}.}
\label{fig2_c}
\end{figure}

The behavior of the entanglement for the pairs of cavities with equal indices as a function of the bandwidth of the driving reservoir is reported in Fig.~\ref{fig2_c}, in the ideal situation in which only the first cavity in each array can dissipate ($\zeta_a \neq 0$) and all other cavities are lossless ($\kappa_j = 0, \forall j$). Here the maximum of the entanglement of the reservoir, namely the value of $\EE_N\al{PO}(0)$ (see Sec.~\ref{Sec.PO}), is fixed by keeping constant the ratio $\alpha/\zeta_b$.
Fig.~\ref{fig2_c} shows how the amount of replicated entanglement reduces with decreasing bandwidth. At large values of $\zeta_b$ the bandwidth is large and the entanglement of the pairs of cavities with equal indices approaches the value of the entanglement of the reservoir. On the other hand, as $\zeta_b$ is lowered, the inter-array pair-entanglement is reduced correspondingly.

\begin{figure}[!t]
\includegraphics[width=8.8cm]{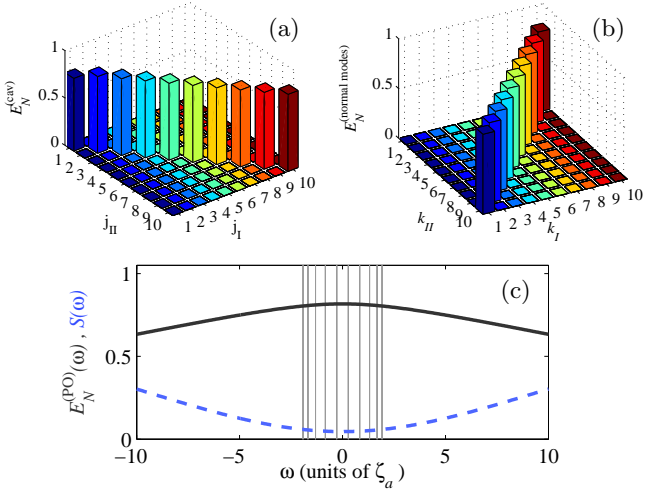}
\caption{(a) Logarithmic negativity for pairs of cavity modes in two equal arrays of $N=10$ cavities: $j_I$ and $j_{II}$ are the indices of the cavities of the first and the second array respectively. (b) Logarithmic negativity for the pairs of normal modes: $k_I$ and $k_{II}$ are indices of the normal modes of the first and the second array respectively, with increasing order in each index corresponding to modes of increasing frequency. (c) Entanglement $E_N\al{PO}\pt{\omega}$ (solid black line) and two-mode squeezing spectrum $S(\omega)$ (dashed blue line) of the output field of the parametric oscillator. The system parameters are $\eta_j=\zeta_a$, $\forall j\in\pg{1,...N}$, $\kappa_j=0$, $\forall j\in\pg{0,...N}$, $\zeta_b=10\zeta_a$ and $\alpha=6.48\zeta_a$ ($\bar\alpha_+=16.48\zeta_a$, $\bar\alpha_-=3.52\zeta_a$). These results correspond to the rightmost arrow in Fig.~\ref{fig2_c}.}
\label{fig2_a}
\end{figure}
\begin{figure}[!h]
\includegraphics[width=8.8cm]{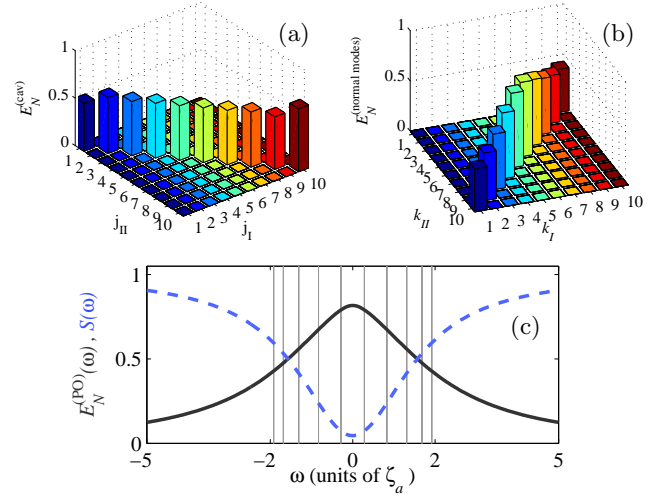}
\caption{As in Fig.~\ref{fig2_a} with $\zeta_b=\zeta_a$ and $\alpha=0.648\zeta_a$. These results correspond to the central arrow in Fig.~\ref{fig2_c}.}
\label{fig2_b}
\end{figure}
\begin{figure}[!h]
\includegraphics[width=8.8cm]{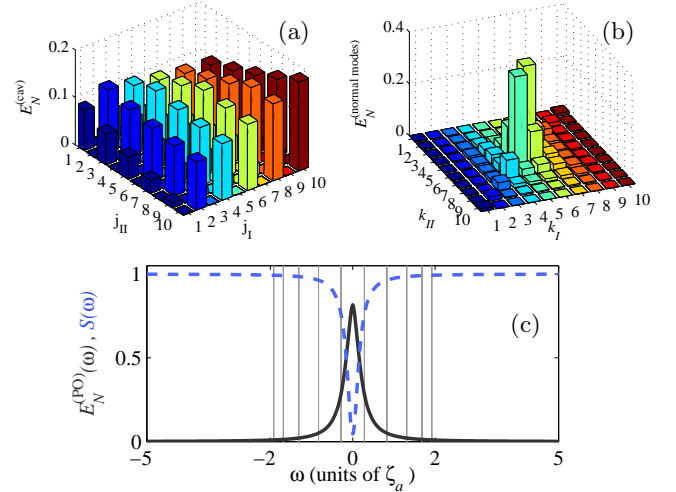}
\caption{As in Fig.~\ref{fig2_a} with $\zeta_b=0.1\zeta_a$ and $\alpha=0.0648\zeta_a$. These results correspond to the leftmost arrow in Fig.~\ref{fig2_c}.}
\label{fig2_d}
\end{figure}

The arrows on the upper part of Fig.~\ref{fig2_c} correspond to the result of Fig.~\ref{fig2_a}, \ref{fig2_b}  and \ref{fig2_d}.
Fig.~\ref{fig2_a} corresponds to the rightmost arrow. Here both $\bar\alpha_+$ and $\bar\alpha_-$ are larger then the parameters of the arrays ($\eta_j$, $\kappa_j$ and $\zeta_a$), the broadband condition is satisfied, and we observe a substantial entanglement replication. Specifically,
the plot in Fig.~\ref{fig2_a} (a) reports the steady-state pairwise entanglement of all the inter-array pairs of cavities (including those with different indices) for two equal arrays of $N=10$ cavities each. We observe that only the cavities with equal indices are strongly entangled, in agreement with the predictions of the broadband model~\cite{Zippilli2013}.
The fact that in this case the squeezing bandwidth of the reservoir is large can be seen looking at Fig.~\ref{fig2_a} (c). Here, the values of the logarithmic negativity $E_N\al{PO}\pt{\omega}=\EE_N\al{PO}(\omega)/(\EE_N\al{PO}(\omega)+1)$ and of the squeezing spectrum $S(\omega)$
of the output field of the parametric oscillator are compared with the range of frequencies, indicated by the vertical lines, that corresponds to the frequencies of the normal modes of the arrays. The latter are evaluated as the imaginary part of the eigenvalues of the matrix $\MM_a$ defined in Eqs.~\rp{Ma_d0} and \rp{Ma_d1}. In detail, this plot shows that the bandwidth of the squeezing is significantly large compared to the normal mode splitting of the arrays and that all the normal modes feel a reservoir with roughly the same level of entanglement. In this case, the Markovian description of Ref.~\cite{Zippilli2013} is accurate.
The corresponding entanglement for pairs of normal modes of the arrays is reported in Fig.~\ref{fig2_a} (b).
Here, the indices $k_{\xi}$, with $\xi\in\pg{I,II}$, indicate the normal modes of the arrays, where $k_{\xi}=1$ and and  $k_{\xi}=10$  are the lowest and higher frequency modes respectively. We observe that a mode of the first array at frequency, say, $\omega_{k_I}$ relative to the frequency of the cavities in array $I$ is entangled with the reciprocal mode of the second array with opposite frequency $\omega_{k_{II}}=\omega_{N-k_I}=-\omega_{k_I}$.
This can be understood by observing that, as discussed in Sec.~\ref{Sec.PO}, the driving two-mode field exhibits a similar structure of entanglement between the different spectral components.
Namely, the spectral component of the first mode of the squeezed reservoir at frequency $\omega$ relative to the frequency of the cavities is entangled with the component of the second mode at relative frequency $-\omega$, and the normal modes of the cavities simply reproduce this same feature.

When the bandwidth of the correlations of the
squeezed reservoir is reduced as in Fig.~\ref{fig2_b}, corresponding to the central arrow in Fig.~\ref{fig2_c}, the replication mechanism is no longer optimal and the entanglement of the pairs of cavities with equal indices is reduced. In terms of the normal modes it corresponds to maximum entanglement between the normal modes whose frequency is closer to the central frequency of the reservoir,
as illustrated in Fig. \ref{fig2_b} (b).
When the bandwidth is further reduced,
as in Fig.~\ref{fig2_d}, which corresponds to the leftmost arrows in Fig.~\ref{fig2_c},
only few normal modes close to resonance with the driving squeezed reservoir (the central modes) become entangled (Fig.~\ref{fig2_d} (b)). As the normal modes can be expressed in terms of the modes of all the cavities, correspondingly the original entanglement of the reservoir is redistributed among all pairs of cavities, at the price of its magnitude being strongly reduced, as illustrated in Fig.~\ref{fig2_d} (a).

\begin{figure}[t!]
\includegraphics[width=8cm]{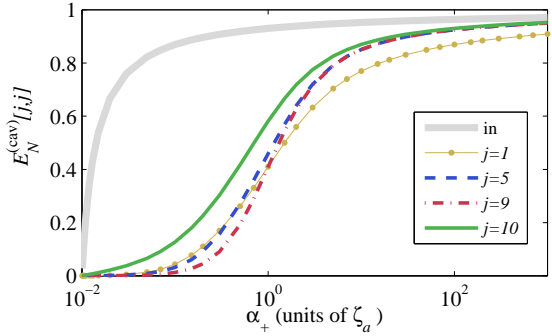}
\caption{Logarithmic negativity for the pairs of cavities with equal indices along two equal arrays of $N=10$ cavities, as a function of the parameter $\bar\alpha_+$, when $\bar\alpha_-=0.01\zeta_a$ is kept constant. The inset specifies the
index of the cavities
corresponding to each curve. The gray line indicates the value of the entanglement at the central frequency of the reservoir. The other parameters are $\eta_j=\zeta_a$, $\forall j\in\pg{1,...N}$ and $\kappa_j=0$, $\forall j\in\pg{0,...N}$.}
\label{fig3}
\end{figure}
\begin{figure}[!t]
\includegraphics[width=8cm]{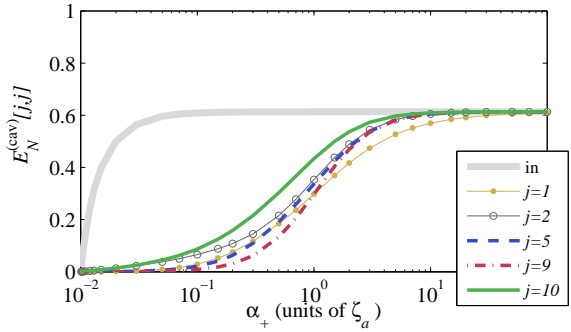}
\caption{
As in Fig.~\ref{fig3} with $\kappa_0=0.5\zeta_b$.
}
\label{fig3b}
\end{figure}

\begin{figure}[t!]
\includegraphics[width=8cm]{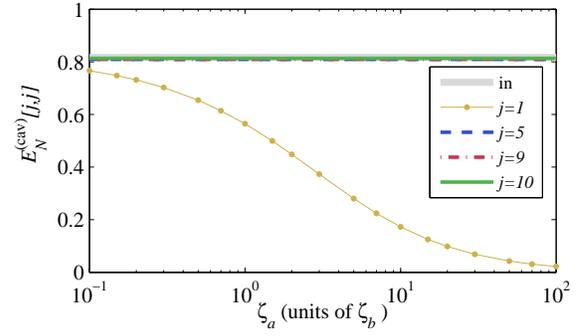}
\caption{Logarithmic negativity for the pairs of cavities with equal indices along two equal arrays of $N=10$ cavities, as a function of the rate $\zeta_a$ at which the array exchange photons with the reservoir. The inset specifies the
index of the cavities
corresponding to each curve. Here all the lines with $j\neq 1$ are almost superimposed. The gray line indicates the value of the entanglement at the central frequency of the reservoir. The other parameters are $\eta_j=0.1\zeta_b$, $\forall j\in\pg{1,...N}$, $\kappa_j=0$, $\forall j\in\pg{0,...N}$,  and $\alpha=0.648\zeta_b$.}
\label{fig4}
\end{figure}
\begin{figure}[!t]
\includegraphics[width=8cm]{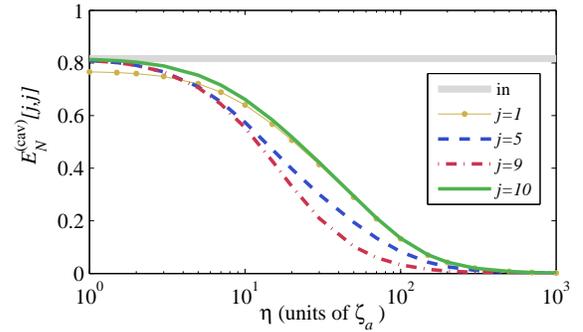}
\caption{Logarithmic negativity for the pairs of cavities with equal indices along two equal arrays of $N=10$ cavities, as a function of the coupling strength $\eta\equiv\eta_j$, $\forall j\in\pg{1,...N}$. The inset specifies the
index of the cavities
corresponding to each curve. The gray line indicates the value of the entanglement at the central frequency of the reservoir. The other parameters are $\forall j\in\pg{1,...N}$, $\kappa_j=0$, $\forall j\in\pg{0,...N}$,  $\zeta_b=10\zeta_a$, and $\alpha=0.648\zeta_b$.
}
\label{fig5_a}
\end{figure}

In Sec.~\ref{Sec.PO}, we have shown that
the correlation functions of the reservoir are characterized by two time scales associated to the parameters $\bar\alpha_\pm$, where $\bar\alpha_+$ determines the  bandwidth of the squeezed quadrature  (and hence of the entanglement) of the field emitted by the parametric oscillator, and the parameter $\bar\alpha_-$ determines the bandwidth of the anti-squeezed quadrature.
Fig.~\ref{fig3} shows the entanglement of pairs of cavity modes as a function of $\bar\alpha_+$ when $\bar\alpha_-$ is fixed to a small value. In this situation, the broadband condition, Eq.~\rp{broad_condition},
is not satisfied. Hence, the Markovian master equation used in~\cite{Zippilli2013} is not valid in the entire range of parameters of Fig.~\ref{fig3}. Nevertheless, we observe that entanglement replication is still in order as long as the squeezing bandwidth $\bar\alpha_+$ is sufficiently large. This result shows that the entanglement replication is not strictly based on the Markovian nature of the dynamics.

A similar situation, with
an additional decay channel at rate $\kappa_0$ affecting the dynamics of the parametric oscillator, is analyzed in Fig.~\ref{fig3b}. It corresponds, for example, to the case in which both mirrors of the Fabry-P\'erot cavity used to realize the parametric oscillator, have a finite transmissivity but only the output from one mirror is efficiently controlled to drive the arrays.
In this case part of the quantum correlations which are built up by the parametric oscillator are lost in the uncontrolled output and the actual reservoir experienced by the arrays is not in a pure squeezed state but in a mixed squeezed thermal state.
In Fig.~\ref{fig3b} we observe the onset of entanglement replication as long as the squeezing bandwidth is of the same order or larger than the characteristic frequencies of the arrays. Specifically, although the maximum entanglement of the reservoir (the thick, gray line in Fig.~\ref{fig3b}) is significantly reduced with respect to the situation illustrated in Fig.~\ref{fig3}, the qualitative behavior of the entanglement is similar and, in the limit of large bandwidth (large $\bar\alpha_+$), the entanglement of the cavities replicates the entanglement of the reservoir. Also in this case we have verified that, when the bandwidth of the reservoir is sufficiently large, {\it only} the modes with equal indices are entangled.

\begin{figure}[!t]
\includegraphics[width=8.8cm]{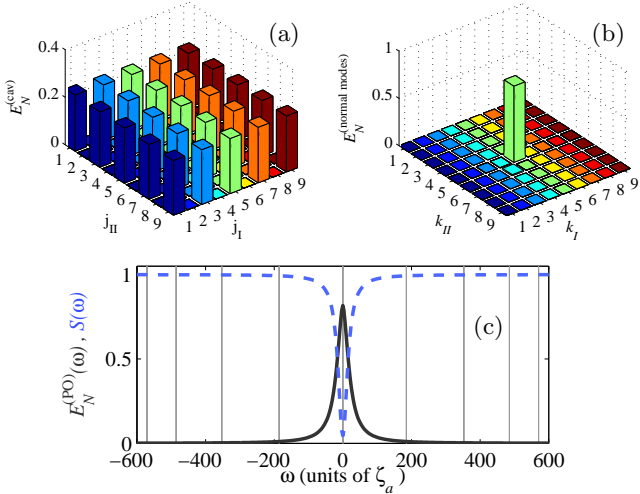}
\caption{As in Fig.~\ref{fig2_a} with $N=9$, $\eta_j=300\zeta_a$, $\forall j\in\pg{1,...N}$, $\kappa_j=0$, $\forall j\in\pg{0,...N}$, $\zeta_b=10\zeta_a$, and $\alpha=6.48\zeta_a$.
}
\label{fig6_a}
\end{figure}

\begin{figure}[!t]
\includegraphics[width=8cm]{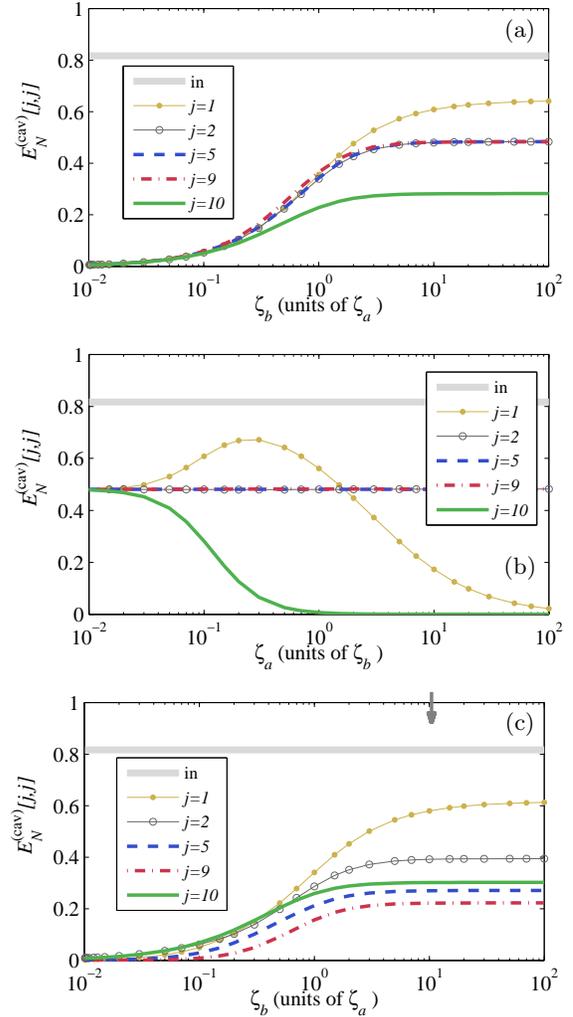}
\caption{
(a) As in Fig.~\ref{fig2_c} with $\kappa_N=\zeta_a$.
(b) As in Fig.~\ref{fig4} with $\kappa_N=\zeta_a$ ($\kappa_N$ is varied together with the parameter $\zeta_a$).
(c) As in Fig.~\ref{fig2_c} with $\kappa_j=0.1\zeta_a$ for $j\in\pg{1,...N}$. The arrow on the top part of plot (c) indicates the parameters corresponding to the results in Fig.~\ref{fig10}.
}
\label{fig7}
\end{figure}
\begin{figure}[!t]
\includegraphics[width=8.8cm]{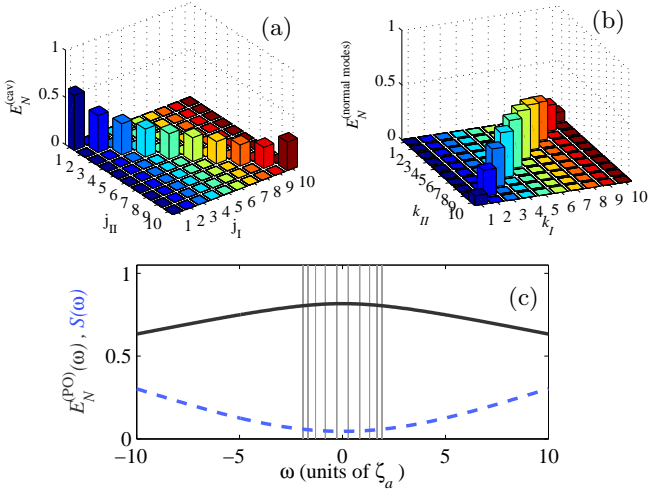}
\caption{As in Fig.~\ref{fig2_a}  with $\zeta_b=\zeta_a$ and $\kappa_j=0.1\zeta_a$ for $j\in\pg{1,...N}$.
These results correspond to the arrow in Fig.~\ref{fig7} (c)}
\label{fig10}
\end{figure}

So far we have analyzed the steady-state entanglement as a function of the parameters of the parametric oscillator. We will now study the steady-state entanglement as a function of the parameters of the arrays when the properties of the parametric oscillator are kept fixed. In particular, we consider the oscillator's parameters given in Fig.~\ref{fig2_a}, corresponding to the situation of good entanglement replication, and we increase either the exchange rate of photons $\zeta_a$ between the driving reservoir and the arrays or the cavity-cavity coupling strengths $\eta_j$ inside each array.
When $\zeta_a$ is increased, as shown in Fig.~\ref{fig4}, only the entanglement of the first pair of cavities is affected, and it decays to zero. Indeed, a large value of $\zeta_a$ means a large line-width of the first cavity in each array. Hence, the system behaves effectively as if the external reservoir would interact directly with the second element in each array. All other pairs of inter-array cavities with equal indices remain unaffected and continue to replicate the entanglement of the reservoir.
On the other hand, when $\zeta_a$ is fixed and the couplings $\eta_j$ are increased, as shown in Fig.~\ref{fig5_a}, then all pairs of cavities are affected; correspondingly, the replicated entanglement decreases for all pairs.

A particular situation is achieved when $\zeta_a$ is sufficiently small and $\eta_j$ is sufficiently large. In this case it is possible to have just a single normal mode resonant with the driving squeezed field. In Fig.~\ref{fig6_a} this is realized using arrays with an odd number of cavities such that there is always a normal mode at zero frequency. Hence when $\eta_j$ is large, the central normal modes of the two arrays are the only modes that interact with the reservoir and they get efficiently entangled, as shown in Fig.~\ref{fig6_a} (b). Due to the particular symmetric situation considered, the central normal modes can be expressed as a linear combination with equal weight of the cavity modes with odd indices. Correspondingly, all the cavity modes with odd indices are entangled in pairs, as shown in Fig.~\ref{fig6_a}  (a), again at the price of a rather small value of the replicated entanglement.

In general, when further sources of dissipation are present, entanglement replication is further degraded, and the pairwise steady-state entanglement can never attain the value of the maximum entanglement of the squeezed reservoir, as shown in Fig.\ref{fig7}. Nevertheless, the qualitative behavior described previously is still observed. In Fig.~\ref{fig7} (a) and (b), we report the situation in which the two arrays are open at both ends with the decay rate of the last cavity of each array equal to the rate at which the first cavity exchanges photons with the driving reservoir: $\kappa_N=\zeta_a$. In this case $E_N\al{cav}[j,j]$ saturates to values smaller than the value of the driving field when the squeezing bandwidth of the reservoir increases. In particular, all but the first and the last pairs saturate to the same value of entanglement, see Fig.~\ref{fig7} (a). It is also interesting to note, see Fig.~\ref{fig7} (b), that in this situation only the entanglement of the first and of the last pairs depends on the dissipation rates of the arrays, while the entanglement for all the other pairs remains constant. The behavior of the first pair at large $\zeta_a$ is similar to that discussed in Fig.~\ref{fig4}, while the decay of $E_N\al{cav}[j,j]$ for the last pair is due to the fact that at large $\kappa_N$ the last pair is effectively decoupled from the rest of the two arrays.

Finally, in Fig.~\ref{fig7} (c) we consider the situation in which all the cavities are lossy. In this case the maximum replicated entanglement is smaller than that of the driving field and it is different for each pair. We note that even if all the cavities dissipate and thus the replicated entanglement is reduced, in any case, as long as the squeezing bandwidth is sufficiently large, {\it only} the inter-array pairs of cavities with the same indices can get entangled, regardless of the effect of additional decay channels affecting the dynamics of the cavities. An example is show in Fig.~\ref{fig10}  which corresponds to the parameters indicated by the arrow in the upper part of Fig.~\ref{fig7} (c).

\section{Experimental implementation}\label{Sec.Exp}

\begin{figure}[!t]
\includegraphics[width=5cm]{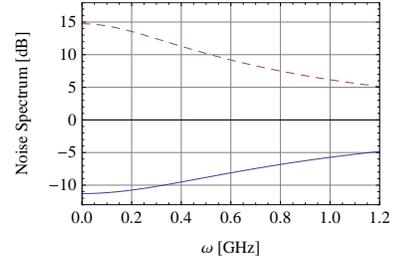}
\caption{Spectrum of the squeezed (solid, blue line) and anti-squeezed (dashed, red line) quadratures, measured in dB and evaluated, using Eqs.~\rp{Sq} and \rp{aSq}, as $10\times\log_{10}[S(\omega)]$ and  $10\times\log_{10}[T(\omega)]$ respectively. The parameters are $\alpha=0.8$~GHz,  $\zeta_b=1.1$~GHz, and $\kappa_0=0.05$~GHz. This behavior is in agreement with the
experimental findings of Ref.~\cite{Ast}.
}
\label{fig8}
\end{figure}

Investigation of the effects of a finite bandwidth is important because the experimental realization of broadband squeezing is extremely challenging. Recently, squeezing of light at telecommunication wavelength with a bandwidth of several GHz and with noise reduction of $\sim3$dB ($\sim10$dB when corrected for detector inefficiency) has been reported in~\cite{Ast}.
Eqs.~\rp{Sq} and \rp{aSq} yield results consistent with those of Ref.~\cite{Ast}, as shown in Fig.~\ref{fig8}, when $\alpha=0.8$~GHz,  $\zeta_b=1.1$~GHz, and $\kappa_0=0.05$~GHz.
\begin{figure}[!t]
\includegraphics[width=8.8cm]{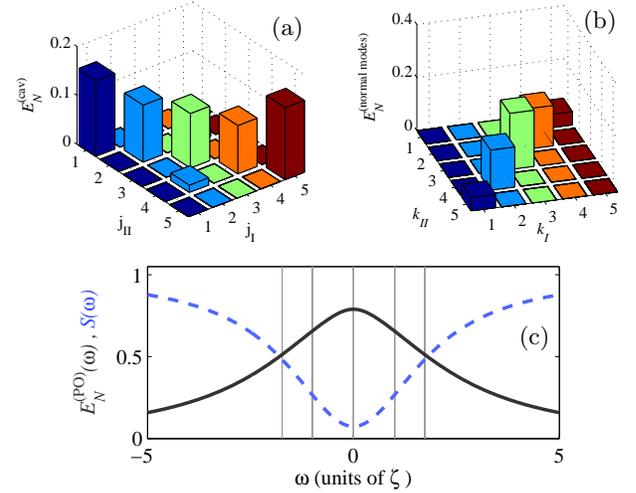}
\caption{As in Fig.~\ref{fig2_a} with $N=5$, $\eta_j=1$~GHz, $\forall j\in\pg{1,...N}$, $\zeta_a=0.1$~GHz $\kappa_j=0.1$~GHz, $\forall j\in\pg{1,...N}$, $\alpha=0.8$~GHz,  $\zeta_b=1.1$~GHz, and $\kappa_0=0.05$~GHz. }
\label{fig9}
\end{figure}

Arrays of cavities in the range of telecom wavelengths are actively investigated in various physical implementations. Realization of cavity arrays looks particularly promising using photonic crystals. Exploiting the latter, in Ref.~\cite{Notomi1,Notomi2,Notomi3} cavity arrays of high quality factor Q ($Q\sim\!\!10^6$) have been studied experimentally: the reported cavity decay rates are of the order of $\sim\!\!1$ GHz, while the cavity-cavity coupling strengths range from 60 to 2000 GHz. They are controlled by realizing different inter-cavity spacings, where larger spacing corresponds to smaller couplings. These parameters, when combined with the squeezing source of Ref.~\cite{Ast}, are surely too large for an experimental verification of the entanglement replication mechanism. Specifically, while it seems reasonable that the coupling strengths can be in principle reduced by increasing the spacing of the arrays using present-day technology~\cite{Notomi1,Notomi2,Notomi3}, significant experimental effort is likely to be required in order to reduce the decay rate of the cavities. On the other hand, photonic-crystal based nano-cavities with larger quality factors are expected to be realizable in the near future~\cite{Taguchi,Minkov}. In order to approach the regime in which entanglement replication is expected to take place, the decay rates should be reduced by roughly one order of magnitude. Employing the range of parameters identified in Fig.~\ref{fig8} for the source of squeezing, we report in Fig.\ref{fig9} the entanglement patterns at the steady state of two arrays, each with $N=5$ cavities, when the decay rates are set at $\kappa_j=0.1$ GHz. From this analysis one sees that entanglement replication takes place and is sufficiently sizeable.

\section{Conclusions}
\label{Conclusions}

We have analyzed the dynamics of non-directly interacting chains of quantum harmonic oscillators when they are locally driven by a common reservoir of entangled particles. In the limit of a broadband reservoir the dynamics of the arrays can be described within the framework of a Markovian master equation. The steady state is then characterized by the formation of many inter-chain pairs of entangled oscillators. Each pair is made of one element of the first array and of the corresponding element, at the same position, of the second array. Ideally, each pair is entangled with the same degree of entanglement of the reservoir, regardless of the length of the arrays and of the position of each pair in the arrays, as originally discussed in Ref.~\cite{Zippilli2013}. In the present work we have investigated how the pairwise entanglement is affected when the reservoir has a generic (finite) bandwidth.
We have considered arrays of linearly coupled optical cavities, and
we have modeled the driving entangled reservoir as the two-mode output field of a non-degenerate optical parametric oscillator. This choice is physically motivated in that it allows to vary the statistical properties of the reservoir by varying experimentally the characteristic parameters of the parametric oscillator, that is the non-linear coupling strength $\alpha$ and the decay rate $\zeta_b$.

As the bandwidth of the squeezing is reduced, the degree of entanglement replication is lowered and the entanglement of the inter-array pairs cannot reproduce the one originally shared by the modes of the reservoir. In all cases, steady-state pairwise entanglement between cavities with the same indices along the two arrays takes place as long as the bandwidth of the squeezing is not smaller than the typical frequency scales of the arrays which are set by the cavity-cavity coupling strengths $\eta_j$ and by the cavity decay rates $\zeta_a$ and $\kappa_j$.

It is also interesting to remark that although the correlations of the operators of the parametric oscillator below threshold are characterized by two time scales, a short time scale $1/(\zeta_b+\alpha)$ related to the squeezed quadrature, and a large time scale $1/(\zeta_b-\alpha)$ related to the anti-squeezed quadrature, only the first one is relevant for the onset of entanglement replication. In particular, even when  $1/(\zeta_b-\alpha)$ is so large to prevent a Markovian description of the dynamics, the entanglement replication can take place as long as $1/(\zeta_b+\alpha)$ is sufficiently small.

In future investigations we plan to study situations involving systems of qubits, either by doping each cavity with a two-level system or by considering spin chains directly driven by an entangled reservoir.

\section*{Acknowledgments}
We thank David Vitali for fruitful discussions. We acknowledge funding by the European Union's Seventh Framework Programme (FP7/2007-2013) under grant agreement number 270843 (iQIT) and grant agreement number 323714 (EQuaM).


\begin{thebibliography}{99}

\bibitem{Amico}
L. Amico, R. Fazio, A. Osterloh, and V. Vedral,
Rev. Mod. Phys. {\bf 80}, 517 (2008).

\bibitem{Cirac97}
J. I. Cirac, P. Zoller, H. J. Kimble, and H. Mabuchi,
Phys. Rev. Lett. {\bf 78}, 3221 (1997).

\bibitem{Kimble}
H. J. Kimble,
Nature {\bf 453}, 1023 (2008).

\bibitem{Zippilli2013}
S. Zippilli, M. Paternostro, G. Adesso, and F. Illuminati, Phys. Rev. Lett. {\bf 110}, 040503 (2013).

\bibitem{Diehl}
S. Diehl, A. Micheli, A. Kantian, B. Kraus, H. P. B\"uchler, and P. Zoller,
Nature Physics {\bf 4}, 878 (2008).

\bibitem{Verstraete}
F. Verstraete, M. M. Wolf, and J. I. Cirac,
Nature Physics {\bf 5}, 633 (2009).

\bibitem{Gardiner93}
C. W. Gardiner,
Phys. Rev. Lett. {\bf 70}, 2269 (1993).

\bibitem{Gardiner94}
C. W. Gardiner and A. S. Parkins,
Phys. Rev. A {\bf 50}, 1792 (1994).

\bibitem{Carmichael93}
H. J. Carmichael,
Phys. Rev. Lett. {\bf 70}, 2273 (1993).

\bibitem{Carmichael94}
P. Kochan and H. J. Carmichael,
Phys. Rev. A {\bf 50}, 1700 (1994).

\bibitem{Neergaard-Nielsen}
J. S. Neergaard-Nielsen, M. Takeuchi, K. Wakui, H. Takahashi, K. Hayasaka, M. Takeoka, and M. Sasaki,
Phys. Rev. Lett. {\bf 105}, 053602 (2010).

\bibitem{Lee}
N. Lee, H. Benichi, Y. Takeno, S. Takeda, J. Webb, E. Huntington, and A. Furusawa,
Science {\bf 332}, 330 (2011).

\bibitem{Steinlechner}
S. Steinlechner, J. Bauchrowitz, M. Meinders, H. M\"uller-Ebhardt, K. Danzmann, and R. Schnabel,
Nature Photonics {\bf 7}, 626 (2013).

\bibitem{Braunstein}
S. L. Braunstein and P. van Loock,
Rev. Mod. Phys. {\bf 77}, 513 (2005).

\bibitem{Weedbrook}
C. Weedbrook, S. Pirandola, R. García-Patrón, N. J. Cerf, T. C. Ralph, J. H. Shapiro, and S. Lloyd,
Rev. Mod. Phys. {\bf 84}, 621 (2012).

\bibitem{Gripp}
J. Gripp, S. L. Mielke, and L. A. Orozco,
Phys. Rev. A {\bf 51}, 4974 (1995).

\bibitem{Stannigel}
K. Stannigel, P. Rabl, and P. Zoller,
New J. Phys. {\bf 14}, 063014 (2012).

\bibitem{Adesso}
G. Adesso and F. Illuminati,
J. Phys. A: Math. Theor. {\bf 40}, 7821 (2007).

\bibitem{Reid1}
M. D. Reid,
Phys. Rev. A {\bf 40}, 913 (1989).

\bibitem{Reid2}
M. D. Reid and P. D. Drummond,
Phys. Rev. A {\bf 40}, 4493 (1989).

\bibitem{Reid3}
P. D. Drummond and M. D. Reid,
Phys. Rev. A {\bf 41},  3930 (1990).

\bibitem{Reid4}
K. Dechoum, P. D. Drummond, S. Chaturvedi, and M. D. Reid,
Phys. Rev. A {\bf 70}, 053807 (2004).

\bibitem{Ast}
S. Ast, M. Mehmet, and R. Schnabel,
Opt. Express {\bf 21}, 13572 (2013).

\bibitem{Notomi1}
M. Notomi, E. Kuramochi, and T. Tanabe,
Nature Photonics {\bf 2}, 741 (2008).

\bibitem{Notomi2}
M. D. Birowosuto, A. Yokoo, H. Taniyama, E. Kuramochi, M. Takiguchi, and M. Notomi,
Journal of Applied Physics {\bf 112}, 113106 (2012).

\bibitem{Notomi3}
M. Notomi, K. Nozaki, A. Shinya, S. Matsuo, and E. Kuramochi,
IEICE Electronics Express {\bf 10}, 20132003 (2013).

\bibitem{Taguchi}
Y. Taguchi, Y. Takahashi, Y. Sato, T. Asano, and S. Noda,
Optics express {\bf 19}, 11916 (2011).

\bibitem{Minkov}
M. Minkov, U. P. Dharanipathy, R. Houdr\'e, and V. Savona,
Opt. Express {\bf 21}, 28233 (2013).

\end{thebibliography}
\end{document}